\documentclass[aps,prc,twocolumn,superscriptaddress,floatfix,10pt]{revtex4-2}
\usepackage{amssymb}
\usepackage{amsmath}
\usepackage{bbm}
\usepackage{dsfont}
\usepackage{braket}
\usepackage{graphicx}
\usepackage{array,multirow}

\usepackage{xcolor}
\usepackage[normalem]{ulem}

\begin{document}
\title{Complex-scaled no-core shell model calculations of bound and unbound nuclear states in light nuclei}

\author{A. T. Kruppa}
\affiliation{HUN-REN Institute for Nuclear Research, Debrecen, Hungary}
\author{N. Michel}
\email{nicolas.michel@impcas.ac.cn}
\affiliation{CAS Key Laboratory of High Precision Nuclear Spectroscopy, Institute of Modern Physics,Chinese Academy of Sciences, Lanzhou 730000, China}
\affiliation{School of Nuclear Science and Technology, University of Chinese Academy of Sciences, Beijing 100049, China}
\author{Xin-le Shang}
\email{shangxinle@impcas.ac.cn}
\affiliation{CAS Key Laboratory of High Precision Nuclear Spectroscopy, Institute of Modern Physics,Chinese Academy of Sciences, Lanzhou 730000, China}
\affiliation{School of Nuclear Science and Technology, University of Chinese Academy of Sciences, Beijing 100049, China} 
\author{Wei Zuo}
\affiliation{CAS Key Laboratory of High Precision Nuclear Spectroscopy, Institute of Modern Physics,Chinese Academy of Sciences, Lanzhou 730000, China}
\affiliation{School of Nuclear Science and Technology, University of Chinese Academy of Sciences, Beijing 100049, China} 
\begin{abstract}
  The complex scaling method is commonly used to describe decaying states, but its applications are limited because the Hamiltonian operator must contain only relative coordinates.
  This has hindered the use of complex scaling in models defined with laboratory single-particle coordinates,
  and in particular one of the most important model in low-energy nuclear physics, the no-core shell model.
  
  We will then present a straightforward procedure for introducing complex scaling in the no-core shell model in order to calculate nuclear resonance states.
	For that matter, the complex-scaled two-body matrix elements must firstly be determined, and the resulting many-body Hamiltonian complex symmetric matrix must be diagonalized afterwards.
  Applications pertain to the bound ground states of the lightest nuclei $^2{\rm H}$, $^3{\rm H}$, $^3{\rm He}$, and $^4{\rm He}$, 
  as well as the resonance ground states of $^5$He and $^5$Li, whereby the realistic interaction Daejeon16 is utilized.
\end{abstract}

\newcommand{\beq}{\begin{equation}}
\newcommand{\eeq}{\end{equation}}

\def\tu{{\tilde u}}
\def\la{\langle}
\def\ra{\rangle}
\def\ord#1{{\cal O}(#1)}
\def\mb#1{{\mathbf{#1}}}
\def\mr#1{{\mathrm{#1}}}
\def\ci{{\rm i}}

\maketitle

\section{Introduction}

{
The position and width of resonances encode essential physical information about the energy at which a nuclear system forms a quasi-bound state and the corresponding decay width. These parameters are fundamental to a detailed description of the continuum spectrum.
In nuclear reaction theory, particularly in models of nucleon capture and resonance scattering, resonance parameters critically determine reaction rates, cross sections, and decay probabilities. Their role is especially pronounced in nuclear astrophysics, where resonant reaction rates are key to nucleosynthesis pathways.
Experimentally, resonance parameters enable direct comparisons with observables such as resonant peaks in cross sections, phase shift variations, and decay spectra, offering stringent tests for theoretical models.
Contemporary nuclear structure and reaction theories increasingly aim for predictive power from ab initio approaches. In this context, accurately computing resonance energies and widths is crucial for validating nuclear interactions and many-body computational methods. These parameters are not peripheral but central to the reliable modeling of the nuclear continuum. They serve as critical benchmarks for theory and play a pivotal role in elucidating the structure and dynamics of weakly bound and unbound nuclear systems. As research extends into the limits of nuclear stability and the continuum frontier, precise resonance characterization becomes indispensable.
}

Ab initio methods aim at describing nuclear systems from first principles, starting from the fundamental interactions between nucleons \cite{epe09}.
They are validated and improved using, e.g., optimization tools \cite{eks13} or Bayesian inference \cite{neu18}. 
Among the most prominent ab initio techniques is the Green’s function Monte Carlo method \cite{pip01}, the coupled cluster method \cite{kum78,hag14}, and the no-core shell model (NCSM) \cite{bar13}.

The application of ab initio methods to explore the continuum spectrum and nuclear reactions represents an important aspect of modern nuclear theory. 
Ab initio methods using Faddeev and Faddeev–Yakubovsky equations \cite{alt67,fad93} have been developed very early to effectively describe nuclear reactions of three and four nucleon systems. 
A breakthrough in ab initio calculations for light nuclei has been the combination of NCSM with resonating group method \cite{nav10},
the NCSM with continuum coupling, deemed as NCSMC \cite{PhysRevC.87.034326} and continuum techniques \cite{nav09,nav16,bar13b}.
The SS-HORSE method, introduced in Refs.\cite{shi16,shi16b,shi18,maz17,maz22,maz24}, offers a valuable extension of the NCSM, by enabling therein the calculation of the S-matrix and thus of resonances states.

{Traditional scattering-based methods depend on asymptotic boundary conditions and the matching of scattering solutions at large distances. The phase shift can be derived from the distribution of the discretized continuum states obtained from a NCSM calculation. 
In contrast to these indirect methods, the Gamow Shell Model (GSM) offers an innovative approach that can directly determine the position and width of a resonant state. The GSM}  extends the traditional shell model by including single-particle basis resonant and continuum states. 
GSM is particularly effective in describing weakly bound and unbound nuclear states, either in core+valence particles or no-core pictures \cite{mic21,pap13,fos17,Li19}.

{
The complex scaling (CS) method is a powerful tool in quantum mechanics for studying resonance states in atomic, molecular, and nuclear systems \cite{rei82,myo14,myo20}. It involves the rotation of particle coordinates into the complex plane to localize diverging wave functions.
CS offers notable advantages over other methods. It allows for a direct extraction of resonance energies and widths from complex eigenvalues, thereby eliminating calculations involving the scattering matrix or phase shift analysis. Mathematically, CS transforms the original Schr{\"o}dinger equation into a bound-state problem with complex boundary conditions.
Numerically, CS simplifies computations by enabling the use of standard $L^2$-integrable bases such as Gaussians or harmonic oscillator (HO) states, so that no matching to asymptotic wave functions intervenes. This makes it especially effective for describing decaying states with techniques developed for bound states.
CS is widely applied in analyzing unstable quantum states involving bosons or fermions \cite{kru90,mas14,kru14,myo23,ods23,oga22,zha22,die22}.
Unlike conventional and continuum shell models that treat bound and unbound states separately \cite{Volya06}, CS unifies these regimes under a single Hamiltonian framework \cite{pap15,yapa2025}.
Additionally, CS facilitates the calculation of scattering amplitudes in nuclear reactions \cite{car14,laz12,myo23,pap15}.
In particular, two recent papers among those cited presented new methods to tackle the challenges of the application of CS to many-body systems.
One paper combined CS, the similarity renormalization group method, and the translationally invariant NCSM \cite{yag25},
while the other incorporated CS into the eigenvector continuation framework to enable accurate bound-to-resonance extrapolations in few-body and many-body systems \cite{yapa2025}.


In traditional shell model calculations, only the valence particles are considered above a closed core; 
in NCSM, all nucleons are active and large HO basis spaces are utilized, with possible extrapolations to infinite model spaces \cite{coo12}.
Two-body matrix elements (TBMEs) are prevalent in shell model calculations.
Added to that, in nuclear physics, suppressing the center-of-mass (c.m.) motion is obligatory, which is achieved by the so-called Lawson correction \cite{law80}.
The main obstacle to the applicability of CS {in shell model} is its strict mathematical background \cite{agu71,bal71,Reed78}, which requires the use of relative coordinates.
We will see that it can be circumvented by exploiting the properties of the HO basis, method deemed in this paper as CS-NCSM.
The CS-NCSM is a direct way to determine the parameters of resonances in a many-body framework.
As a first application of CS-NCSM, we will calculate well-bound and fairly broad resonance ground states states of $A=2-5$ nuclei using the realistic Daejeon16 interaction \cite{shi16}.

The article is structured as follows. All applications of the CS method utilize the CS intrinsic Hamiltonian formalism, which is briefly overviewed in Section 2.
In Section 3, we present our new method that demonstrates the application of CS-NCSM, including the Lawson correction, while emphasizing its practical interest.
Section 4 outlines the computational approach and issues for calculating TBMEs when using CS.
Finally, in Section 5, we provide our numerical results, beginning with an analysis of the ground states of the nuclei 
$^2{\rm H}$, $^3{\rm H}$, $^3{\rm He}$ and $^4{\rm He}$. We then consider the resonance ground states of $^5$He and $^5$Li. The conclusion of the paper is found in Section 6.

\section{Complex scaled intrinsic Hamiltonian}

We will derive the formulation of the complex-scaled intrinsic Hamiltonian in detail.
Our approach closely follows that of Ref.\cite{Reed78}.
We will also explicit utilize internal coordinates in order to demonstrate that the center of mass kinetic energy operator commutes with the complex-scaled intrinsic operator.

The unitary dilatation operator \( u_\theta \) is defined in \( L^2(\mathbb{R}^3) \) by the formula :
\begin{equation}\label{U_dilation}
(u_\theta \Psi)({\bf r}) = e^{\frac{3}{2}\theta}~\Psi(e^{\theta} {\bf r}),
\end{equation}
where \( \theta \) is a real number. The Laplace operator \( \Delta_{\bf r} \) simply transforms under $u_\theta$, as it satisfies the relation : 
$u_\theta \Delta_{\bf r} u_\theta^{-1} = e^{-2\theta} \Delta_{\bf r}.$
Furthermore, the transformed Laplace operator has an analytic continuation to the complex plane.
In the context of the CS method, the transformed interaction is constrained so that this analytic continuation holds for the entire Hamiltonian.

Let us define the complex-scaled kinetic operator from its initial definition ${\tilde H}_0=\sum_{i=1}^A(2m_i)^{-1}\Delta_{{\bf r}_i}$, with $m_i$ the mass of the $i$-th nucleon :
\begin{equation}\label{cskin}
{\tilde H}_0(\theta)={\rm e}^{-2\theta}{\tilde H}_0.
\end{equation}
The c.m.~coordinate is denoted by $\bf R$ and the $N-1$ nucleon coordinates are defined by ${\boldsymbol{\eta}}_i={\bf r}_i-{\bf r}_A$ for $i=1,\ldots,A-1$.
This coordinate system provides with a tensor decomposition of the Hilbert space $L^2(\mathbbm{R}^{3A})=L^2(\mathbbm{R})\otimes L^2(\mathbbm{R}^{3A-3})$.
We can then write the operator ${\tilde H}_0$ with tensor products in c.m.~and intrinsic spaces:
\begin{equation}\label{ttcom}
{\tilde H}_0=T_{\rm cm}\otimes \mathbbm{1}+\mathbbm{1}\otimes T_{\rm int}.
\end{equation} 
The c.m.~kinetic energy operator reads :
\begin{equation}\label{tcm}
T_{\rm cm} = -\frac{\Delta_{{\bf R}}}{2 m_A}
\end{equation} 
where $m_A$ is the mass of the nucleus and the intrinsic part of the kinetic energy operator becomes : 
\begin{equation}\label{tkin}
T_{\rm int}=-\sum_{i=1}^{A-1}\frac{\hbar^2}{2m_{iA}}\Delta_{{\boldsymbol{\eta}}_i}+\sum_{i<j}^{A-1}\frac{\hbar^2}{2m_A}\nabla_{{\boldsymbol{\eta}}_i}\nabla_{{\boldsymbol{\eta}}_j},
\end{equation}
where ${m_{iA}}^{-1}={m_i}^{-1}+{m_A}^{-1}$.
From Eqs.(\ref{cskin},\ref{ttcom}), the tensor product decomposition of ${\tilde H}_0(\theta)$ is obtained :
\begin{equation}\label{tkint}
{\tilde H}_0(\theta)=e^{-2\theta}T_{\rm cm}\otimes \mathbbm{1}+\mathbbm{1}\otimes e^{-2\theta}T_{\rm int}.
\end{equation}

We assume that the interaction part of the Hamiltonian is of two-body character :
\begin{equation}
V_{\rm int}=\sum_{i<j}^A V_{ij}(|\mathbf{r}_i-\mathbf{r}_j|).
\end{equation}
If potentials are central and satisfy the very general conditions of Ref.\cite{Reed78} (see Example 1 in p.~185 in Ref.\cite{Reed78}),
$V_{\rm int}$ takes the form of a multiplication operator
possessing an analytic extension in the complex strip $B_\alpha=\{z\in\mathbb{C}\ \vert\ \vert{\rm arg}(z)\vert <\alpha\}$, where $\alpha>0$ :
\begin{equation}\label{vint} 
V_{\rm int}^\theta=\sum_{i<j}^AV_{ij}(e^{\theta}|\mathbf{r}_i-\mathbf{r}_j|)
\end{equation}
For $\theta\in B_\alpha$ complex, the analytically extended interaction is also a multiplication operator, so that Eq.(\ref{vint}) is valid in this situation as well.
We can express $V_{\rm int}^\theta$ from the $A-1$ coordinates ${\boldsymbol{\eta}}_i$ only :
\begin{equation}\label{vinteta}
V_{\rm int}^\theta=\sum_{i=1}^{A-1}V_{iA}(e^{i\theta}\vert{\boldsymbol{\eta}}_i\vert)
+\sum_{i<j}^{A-1}V_{ij}(e^{i\theta}\vert{\boldsymbol{\eta}}_i-{\boldsymbol{\eta}}_j\vert).
\end{equation}
The general definition of a dilation analytic potential is, however, more involved (see the definition in p.~184 of Ref.\cite{Reed78}).

The complex-scaled nuclear Hamiltonian $\tilde H(\theta) = {\tilde H}_0(\theta) + V_{\rm int}^\theta$ thus writes :
\begin{equation}\label{fulth}
{\tilde H}(\theta)=e^{-2\theta}T_{\rm cm}\otimes \mathbbm{1}+\mathbbm{1}\otimes \left (e^{-2\theta}T_{\rm int}+V_{\rm int}^\theta\right).
\end{equation}  
Let $H_{\rm int}^\theta$ denote ${\tilde H}(\theta)$ with its c.m.~motion removed (see definition p.~190 of Ref.\cite{Reed78}).
According to Eqs.(\ref{tkin},\ref{tkint},\ref{vint},\ref{fulth}), the complex-scaled expression of $H_{\rm int}^\theta$ reads :
\begin{eqnarray}\label{hintt}
&&H_{\rm int}^\theta=e^{2\theta}\left(-\sum_{i=1}^{A-1}\frac{\hbar^2}{2m_{iA}}\Delta_{{\boldsymbol{\eta}}_i}+\sum_{i<j}^{A-1}\frac{1}{m_A}\nabla_{{\boldsymbol{\eta}}_i}\nabla_{{\boldsymbol{\eta}}_j}\right)\nonumber\\
&+&\sum_{i=1}^{A-1}V_{iA}(e^{\theta}\vert{\boldsymbol{\eta}}_i\vert)\nonumber\\
&+&\sum_{i<j}^{A-1}V_{ij}(e^{\theta}\vert{\boldsymbol{\eta}}_i-{\boldsymbol{\eta}}_j\vert)\ \ \  \theta\in B_\alpha.
\end{eqnarray}

Since the \( A \) c.m.~and intrinsic coordinates, respectively equal to \( \mathbf{R} \) and \( \boldsymbol{\eta}_i \) for \( i = 1, \ldots, A-1 \), are independent, 
the center-of-mass kinetic energy operator (see Eq.(\ref{tcm})), and the complex-scaled intrinsic operator (see Eq.(\ref{hintt})), commute.
This conclusion remains valid for every set of intrinsic coordinates, such as Jacobi coordinates, which differ from \( \boldsymbol{\eta}_i \) in general \cite{yapa2025}.

It is important to note that the standard intrinsic Hamiltonian \( H_{\rm int} \), as used in scenarios pertaining to NCSM,
can be derived from Eq.(\ref{hintt}) when \(\theta = 0\) : \( H_{\rm int} = H_{\rm int}^{\theta=0} \).
In numerical calculations, it is customary to choose a pure imaginary value for \(\theta\) in Eq.(\ref{hintt}), as coordinate length scaling is unimportant for resonances.
For the remainder of this paper, we then adopt the notation \( i\theta \), where \( 0 < \theta < \alpha \), to represent the initial \(\theta\) value of Eq.(\ref{hintt}).

$H_{\textrm{int}}^\theta$ possesses remarkable properties in the complex plane.
Indeed, according to the so-called Aguilar-Balslev-Combes (ABC) theorem \cite{agu71,bal71} (see also Ref.\cite{Reed78}),
the transformed Hamiltonian $H_{\textrm{int}}^\theta$ has the same discrete bound eigenspectrum as $H_{\textrm{int}}$,
whereas its continuous spectrum is rotated by an angle $\theta$ in the complex plane.
Added to that, resonance eigenstates appear in the eigenspectrum, thereby represented by square-integrable many-body wave functions \cite{agu71,bal71}.
All CS applications are restricted to the use of intrinsic Hamiltonians due to the ABC theorem \cite{agu71,bal71}.
Figure \ref{H_CS} presents a typical schematic spectrum of an intrinsic CS Hamiltonian.

Let us comment on the Coulomb potential, whose infinite range always demands care.
The two-body Coulomb force is entirely characterized by the form factor \( \frac{1}{| \mathbf{r} |} \), so that it is dilation analytic by construction.
This allows for the direct application of the complex scaling method based on the ABC theorem (see Example 1 on p.~185 of Ref.\cite{Reed78} and Ref.\cite{Sim73}).
Conversely, the Coulomb potential arising from a uniformly charged sphere, as widely applied in coupled-channel nuclear models, is not dilation analytic.
This caveat can be circumvented, for example, by introducing a charged sphere generated by a proton distribution of Gaussian type \cite{Saito69,Myo98,idb08}.

It is important to note that dilation analytic potentials do not necessarily have to be local multiplication operators;
they can also be represented as separable interactions in the context of nuclear physics (see Example 2 on p.~186 in Ref.\cite{Reed78}).

\begin{figure}
\caption{{The schematic spectrum of the CS intrinsic Hamiltonian. It is assumed that the intrinsic Hamiltonian has only  one bound state, one resonance state and two thresholds. 
The bound and resonance states are displayed with black and red circles. The thresholds are indicated by green circles, and thick, solid lines sign the rotated-down continuums. The dotted circular arc designates an angular value of $2\theta$.\label{H_CS}}} 
\includegraphics[scale=0.7]{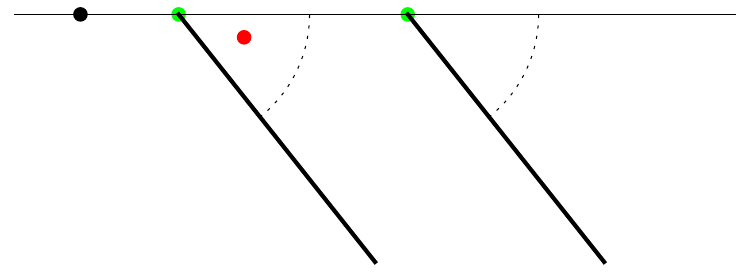}
\end{figure}

\section{Complex-scaled no-core shell model}
To solve the many-body problem, we utilize Slater determinants constructed from spherical HO single-particle orbitals. 
These determinants are solutions of the eigenvalue problem of the independent-particle HO Hamiltonian:
\beq\label{hoop}
H_{\textrm{HO}} = \sum_i \left( \frac{\mathbf{p}_i^{\,2}}{2m} + \frac{1}{2} m \omega^2 \mathbf{r}^2_i \right),
\eeq
where ${\bf p}_i=-i\hbar\nabla_{{\bf r}_i}$.
The non-interacting many-body HO Hamiltonian of Eq.(\ref{hoop}) then writes :
\beq\label{hosep}
H_{\textrm{HO}} = H^{\textrm{HO}}_{\textrm{c.m.}} + H_{\textrm{int}}^{\textrm{HO}},
\eeq
where
\beq\label{horel}
H_{\textrm{int}}^{\textrm{HO}}=T_{\textrm{int}}+\frac{1}{A} \sum_{i<j} \frac{1}{2}m\omega^2(\mathbf{r}_i-\mathbf{r}_j)^2,
\eeq
where the c.m.~wave function is demanded to be the ground state of the HO-c.m.~potential:
\beq\label{hocm}
H_{\textrm{c.m.}}^{\textrm{HO}} = \frac{\mathbf{P}^2}{2M}+\frac{1}{2}M\omega^2\mathbf{R}^2,
\eeq
where ${\bf P}=-i\hbar\nabla_{{\bf R}}$.

The intrinsic Hamiltonian $H_{\rm int }$  becomes a matrix when it is represented by a basis of Slater determinants constructed from HO single-particle orbitals.
As Slater determinants are defined with laboratory coordinates, the nucleus c.m.~motion could unphysically mix with the intrinsic degrees of freedom of nucleons, which are solely responsible of nuclear binding.

The c.m.~spurious motion is properly suppressed if the eigenstates of $\tilde H=\tilde H_0+V_{\rm int}$ separate in c.m.~and relative parts \cite{law80} :
\begin{equation}\label{HO_CM_int}
\ket{\Psi} = \ket{\Psi_{\textrm{c.m.}}} \otimes \ket{\Psi_{\textrm{int}}},
\end{equation}
where $\ket{\Psi}$ ($\ket{\Psi_{\textrm{int}}}$) is a $\tilde H$($H_{\rm int}$) eigenstate and $\ket{\Psi_{\textrm{c.m.}}}$ is an eigenfunction of the c.m.~HO Hamiltonian,
denoted by $H_{\textrm{c.m.}}^{\textrm{HO}}$ \cite{law80} :
\begin{equation}\label{HO_CM}
H_{\textrm{c.m.}}^{\textrm{HO}}\ket{\Psi_{\textrm{c.m.}}} = E_{\textrm{c.m.}}\ket{\Psi_{\textrm{c.m.}}}.
\end{equation}
The factorization of Eq.(\ref{HO_CM_int}) is performed {in NCSM} by diagonalizing the following operator along the so-called Lawson method \cite{law80} :
\begin{equation}\label{nocore}
H_{{\omega,}\beta} = H_{\textrm{int}} + \beta \left( H_{\textrm{c.m.}}^{\textrm{HO}} - \frac{3}{2} \hbar \omega \right),
\end{equation}
where $\beta$ is a large number.
All many-body basis states of HO energy smaller or equal to $N \hbar \omega$ must be included in the NCSM space for Eq.(\ref{HO_CM_int}) to be imposed \cite{law80}.
This type of basis space is deemed as an $N \hbar \omega$ space, this denomination being standard \cite{cau05}. The factorization of the c.m.~and the intrinsic state is preserved in the truncated space \cite{her16}.

The ABC theorem can be applied to $H_{\textrm{int}}$ in Eq.(\ref{nocore}).
$H_{\textrm{CM}}^{\textrm{HO}}$ in Eq.(\ref{nocore}) is impervious to that treatment, however, because it is defined with c.m.~coordinates.
Apparently, this would prevent the ABC theorem to be of interest in NCSM.
This caveat can be solved by noticing that the $\ket{\Psi_{\textrm{c.m.}}}$ HO c.m.~ground state of $H^{\textrm{HO}}_{\textrm{c.m.}}$ (see Eq.(\ref{HO_CM_int})) does not have to be complex-scaled.
We wherefore define the CS-NCSM Lawson-corrected Hamiltonian operator from Eq.(\ref{nocore}) by replacing $H_{\textrm{int}}$ by $H^\theta_{\textrm{int}}$ therein, while its c.m.~part remains the same :
{\begin{equation}\label{full}
H^\theta_{\omega,\beta}=H_{\textrm{int}}^\theta+\beta\left(H_{\rm c.m.}^{\rm HO}-\frac{3}{2}\hbar\omega\right).
\end{equation}}

Since the two operator terms on the right-hand side of Eq.(\ref{full}) act on different coordinate systems, the exact solution of the eigenvalue problem of $H^\theta_{\omega,\beta}$ reads:
\begin{equation}\label{decomp}
  \ket{\Psi^\theta} = \ket{\Psi_{\rm c.m.}} \otimes \ket{\Psi^\theta_{\textrm{int}}},
\end{equation}
where :
\begin{equation}\label{H_theta}
  H^\theta_{\omega,\beta} \ket{\Psi^\theta} = E^\theta \ket{\Psi^\theta},
\end{equation}
and
\begin{equation}\label{H_theta_int}
  H^\theta_{\textrm{int}} \ket{\Psi^\theta_{\textrm{int}}} = E^\theta_{\textrm{int}} \ket{\Psi^\theta_{\textrm{int}}},
\end{equation}
whereby $E^\theta = E^\theta_{\textrm{int}} + \beta ( E_{\textrm{c.m.}} - (3/2) \hbar \omega)$ befalls from Eq.(\ref{full}).
It follows from Eq.(\ref{decomp}) that c.m.~motion is completely decoupled from the intrinsic wave function in the exact solution, therefore implying that spurious c.m.~motion wanes out. 

The intrinsic complex-scaled operator of Eq.(\ref{hintt}) and the HO center-of-mass operator of Eq.(\ref{hocm}) commute because they rely on different sets of dynamical coordinates.
Let us the consider the operator $P$ which projects onto a subspace of Slater determinants constructed from HO single-particle states,
in which approximate Hamiltonian eigenstates are decomposed.

Factorization of the center-of-mass and intrinsic wave functions within that truncated model space can be achieved
if the operators $P H^\theta_{\text{int}} P$ and $P H_{\text{c.m.}}^{\text{HO}} P$ commute, as this indicates that they possess common eigenfunctions.
For that matter, it is sufficient to generalize the proof of the separation of intrinsic and center-of-mass motion of Ref.\cite{her16} to the complex-scaled case.
The premise leading to that demonstration is that $P$, which projects into the $ N \hbar \omega$ model space, verifies : $[P,H_{\text{c.m.}}^{\text{HO}}] = 0$.
In fact, generalization to the complex-scaled case is straightforward because the proof of Ref.\cite{her16} does not depend on the $\theta$ value associated with the complex-scaled intrinsic operator.

Figure \ref{H_CS_CM} exhibits the problematic of c.m.~spuriosity in CS Hamiltonians.
A CS-NCSM numerical calculation can then be carried out as in a standard NCSM calculation.
The interaction potential has to be complex-scaled in CS-NCSM when the TBMEs are calculated, which will be the topic of the next section. 
Otherwise, the Hamiltonian matrix becomes complex symmetric instead of Hermitian, whose diagonalization procedure is based on the variational principle generalized to complex energies \cite{Moy11,mic21}.

\begin{figure}
{\caption{Schematic spectrum of the CS Lawson-corrected Hamiltonian. The intrinsic Hamiltonian only has two thresholds. Subsystems do not have resonances and the system have one bound and one resonance state. The bound and resonance states are represented by black and red circles. The thresholds are denoted by green circles, and the rotated down continuums are represented by thick solid lines. Only the part of the spectrum is shown where the maximum excitation of the c.m.~motion is $2\hbar \omega$. The dotted circular arc represents an angle of $2\theta$. \label{H_CS_CM} }}
\includegraphics[scale=0.7]{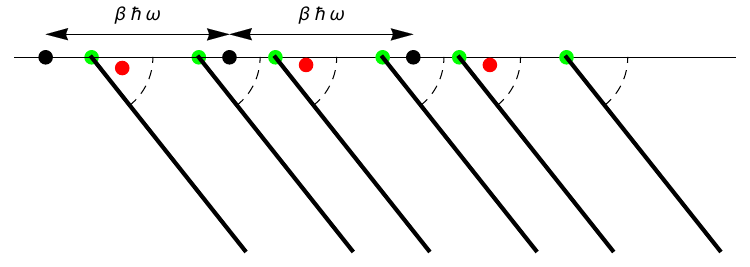}
\end{figure}

The explicit form of the CS Lawson-corrected Hamiltonian is hereby provided.
In practical calculations, $H_{\rm c.m.~}^{\textrm{HO}}$ is defined by way of one and two-body operators:
\beq\label{hcm}
H_{\textrm{c.m.}}^{\textrm{HO}}=H_{\textrm{HO}}-H_{\textrm{int}}^{\textrm{HO}},
\eeq
so that we obtain :
\begin{eqnarray}\label{nocore1}
H^\theta_\beta &=& e^{-2i\theta} T_{\textrm{int}} + \sum_{i<j}^A V_{ij} (e^{i\theta} |\mathbf{r}_i - \mathbf{r}_j|) \nonumber \\
&+& \beta \left(H_{\textrm{HO}} - H_{\textrm{int}}^{\textrm{HO}} - \frac{3}{2} \hbar \omega \right).
\end{eqnarray}
The term proportional to $\beta$ in Eq.(\ref{nocore1}) is deemed as the Lawson correction, as it is this operator which relieves eigenstates from spurious c.m.~motion. 
The explicit form of the CS Hamiltonian with Lawson correction writes:
\begin{eqnarray}\label{csnocore2}
H^\theta_{\omega,\beta} &=& e^{-2i\theta}\frac{1}{A} \sum_{i<j} \frac{(\mathbf{p}_i-\mathbf{p}_j)^2}{2m} + \sum_{i<j}^A V_{i,j} (e^{i\theta} |\mathbf{r}_i - \mathbf{r}_j|) \nonumber \\
&+& \beta \sum_i \left(\frac{\mathbf{p}_i^{\,2}}{2m} + \frac{1}{2}m\omega^2\mathbf{r}^2_i\right) \nonumber \\
&-& \frac{\beta}{A} \left( \sum_{i<j} \frac{(\mathbf{p}_i-\mathbf{p}_j)^2}{2m} + \sum_{i<j} \frac{1}{2} m \omega^2 (\mathbf{r}_i - \mathbf{r}_j)^2 \right) \nonumber \\ 
&-& \frac{3}{2} \beta \hbar \omega .
\end{eqnarray}
The operator above not only contains the CS original interaction, but also a additional interaction of two-body type induced by the Lawson correction.

Let us now ponder on the overall numerical cost of CS-NCSM.
Even though NCSM and CS-NCSM possess very different physical content, the computational burden happens to be virtually the same in both situations.
Indeed, the model space dimension of  $ N \hbar \omega$ model spaces is identical in both NCSM and CS-NCSM for a fixed value of $N$.
As the CS-NCSM matrix does not have to be stored in on-the-fly calculations \cite{cau05}, the additional memory storage borne by complex Hamiltonian matrix elements only concerns TBMEs,
whose numerical treatment is minute compared to matrix-times-vector operations (see next section for details).
In fact, the only difference is that a complex symmetric matrix in CS-NCSM must be diagonalized instead of a real symmetric matrix in NCSM.
This demands to store shell model vectors of complex numbers instead of real numbers, while it also generates twice longer calculations.
However, these latter costs can be deemed as negligible compared to the combinatorial increase of model space dimension with the number of valence nucleons in shell model in general \cite{cau05}.

As a consequence, the current computational limit of NCSM for bound states, i.e.~nuclei bearing $A \sim 20$ nucleons, is the same in CS-NCSM,
so that we can expect to able to study these light nuclei with CS-NCSM as well.
This constitutes the fundamental advantage of the CS-NCSM compared to the other numerical methods including continuum coupling in many-body wave functions.
Indeed, NCGSM is much more expensive computationally because of the use of the Berggren basis, whereby each partial wave is discretized with 30 shells typically \cite{mic21}.
Hence, NCGSM can hardly be applied in nuclei bearing $A=6-7$ nucleons or more.
While approaches based on NCSM with resonating group method or NCSMC make use of the HO basis \cite{nav10,PhysRevC.87.034326,nav09,nav16,bar13b},
they demand to solve reaction coupled-channel equations in relative coordinates built from HO NCSM eigenstates,
which is numerically expensive, especially when few-body clusters enter reaction channels.

\section{Complex-scaled two-body matrix elements}

A novel step in CS-NCSM calculations is to determine the TBMEs arising from the CS interaction potential.
These TBMEs can be derived by considering Eqs.(\ref{nocore1},\ref{csnocore2}) with two-body wave functions.
{
  In this section, we describe two methods: one demands the interaction to be known in coordinate space, while the other is based on a separable approximation of the potential.
  Both methods make use of the Talmi-Brody-Moshinsky method \cite{law80,miy23}.

In a CS-NCSM calculation, the CS-TBMEs are given by the equation:
\begin{equation}\label{tbmecs}
V_{\theta}(a,b,c,d;JT) = \braket{a,b;JT | V({\bf r} e^{i\theta}) | c,d;JT}.
\end{equation}
The shorthand notation $a$ represents the three quantum numbers $n_a,l_a$, and $j_a$ (same for $b,c,d$),
whereas $V({\bf r} e^{i\theta})$ is the CS interaction in coordinate space function of the relative position vector ${\bf r}={\bf r}_1-{\bf r}_2$.
The notation $ \ket{a,b; JT}$ refers to angular and isospin coupled HO two-body wave functions composed of single particle HO states $a$ and $b$.
TBMEs are independent of angular momentum and isospin projections, so that they are not mentioned. 
The inner products in Eq.(\ref{tbmecs}) are not complex-conjugated according to the theory of the CS \cite{Moy11}.
If we apply the Talmi-Brody-Moshinsky method to Eq.(\ref{tbmecs}), we obtain the following form for the CS-TBME :
\begin{eqnarray}\label{tbmeint1}
V_{\theta}(a,b,c,d;JT) &=& \sum_{NLSS' j T} \sum_{nl n' l'} \nonumber \\
&& M(NL,nl,j,S,T) \nonumber \\
&\times& M(NL,n'l',j,S',T) \nonumber \\
&\times& V_{nl,n'l',j}^{S,S',T}(\theta),
\end{eqnarray}
where the explicit form of the expression $M(NL,nl,j,S,T)$ is given for example in \cite{law80,miy23}.
The quantum numbers introduced by the  Talmi-Brody-Moshinsky transformation are c.m.~radial quantum
number ($N$), c.m.~orbital angular momentum ($L$), relative radial
quantum numbers ($n,n'$), relative orbital angular momentums ($l,l'$), total
spin ($S,S'$), isospin ($T$) and total angular momentum of the relative motion ($j$) of the two particles.
We use the following notation: 
\begin{equation}\label{relho}
R_{nl}(r)\Big[ Y_l(\hat{r})\otimes \chi_{S}^{(12)}\Big]_{j}\tau_{T}^{(12)} = R_{nl}(r)\Gamma_{ljST}(\hat{r}),
\end{equation}
where $r = |{\bf r}|$ and $\hat r = {\bf r}/r$.
In Eq.(\ref {relho}), $R_{nl}(r)$ is the radial HO wave function, whereas the functions $\chi_{S}^{(12)}$ and $\tau_{T}^{(12)}$ represent the spin and isospin states of two nucleons.

The radial form factor of the interaction determines $V_{nl,n'l',j}^{S,S',T}(\theta)$ in Eq.(\ref{tbmeint1}), as it can be obtained by the integral : 
\begin{equation}\label{integral}
V_{nl,n'l',j}^{S,S',T}(\theta) = \braket{R_{nl} \Gamma_{ljST} | V(re^{i\theta}) |  R_{n'l'} \Gamma_{l'jS'T}}.
\end{equation}
However, Eq.(\ref{integral}) demands $V(re^{i\theta})$ to be explicitly known in coordinate space, which is usually not the case with realistic interactions.
An alternative method has then been devised, which is based on a procedure developed in GSM calculations.
This approach is based on a separable expansion of the potential $V$, denoted as $\tilde{V}$ \cite{mic21,hag06}:
\begin{eqnarray}\label{seppot}
&&\tilde{V} = \sum_T \sum_{\alpha,S} \sum_{\beta,S'} \nonumber \\
&& \ket{R_{n_\alpha l_\alpha} \Gamma_{l_\alpha,j_\alpha,S,T}}~V^{S,S',T}_{\alpha,\beta}~\bra{R_{n_\beta,l_\beta}\Gamma_{l_\beta,j_\beta,S',T}},
\end{eqnarray}
where :
\begin{equation}\label{potmat}
V^{S,S',T}_{\alpha,\beta} = \braket{R_{n_\alpha,l_\alpha}\Gamma_{l_\alpha,j_\alpha,ST} | V | R_{n_\beta,l_\beta}\Gamma_{l_\beta,j_\beta,S'T}}.
\end{equation}
For a given $l_\alpha$ and $l_\beta$ values all possible $j_\alpha$ and $j_\beta$ values are taken into account in Eq.(\ref{seppot}). The upper limits in Eq.(\ref{seppot}) are determined by two quantum numbers 
$n_{max}$ and $l_{max}$ in such a way that $0\le n_\alpha,n_\beta\le n_{max}$ and  $0\le l_\alpha,l_\beta\le l_{max}$. 
While the complex-scaled form of the interaction is used in Eq.(\ref{integral}), this operator is defined by
$U_{\theta}\tilde{V} U_{\theta}^{-1}$ in Eq.(\ref{potmat}),
where the operator $U_{\theta}$ scales only the radial coordinate $r$ of $R_{n_\alpha l_\alpha}(r)$ and $R_{n_\beta l_\beta}(r)$ HO relative states with the rotation angle $\theta$.

The TBME of Eq.(\ref{integral}) in the separable approximation case thus writes:
\begin{eqnarray} \label{cstbme3}
\!\!\!\!\!\!\!\! V_{nl,n'l',j}^{S,S',T}(\theta) &=& \braket{R_{nl} \Gamma_{ljST} | \tilde{V}_\theta | R_{n'l'} \Gamma_{l'jS'T}} \nonumber \\
&=& \sum_T \sum_{\alpha,S} \sum_{\beta,S'} \nonumber \\
&& \braket{ R_{nl} (r e^{-i\theta}) \Gamma_{ljST} | R_{n_\alpha l_\alpha} \Gamma_{l_\alpha,j_\alpha,S,T}}  \nonumber \\
&\times& V^{S,S',T}_{\alpha,\beta}  \nonumber \\
&\times& \braket{R_{n_\beta l_\beta} \Gamma_{l_\beta j_\beta S'T} | R_{n'l'}(r e^{-i\theta})\Gamma_{l'jS'T}},
\end{eqnarray}
where the abreviation $\alpha$ and $\beta$ denote the one-body quantum numbers $n_\alpha,l_\alpha,j_\alpha$ and  $n_\beta,l_\beta,j_\beta$, respectively, and where no complex conjugation acts on $r e^{i \theta}$ \cite{agu71,bal71,Moy11}.
Note that the HO length defining the HO states $R_{n_\alpha l_\alpha}\Gamma_{l_\alpha,j_\alpha,S,T}$ and $R_{n_\beta l_\beta}\Gamma_{l_\beta,j_\beta,S',T}$ in the separable expansion in Eq.(\ref{seppot}) can be different from that of the size parameter of the HO states entering the TBME of Eq.(\ref{tbmecs}).

The method pertaining to the separable expansion above allows to evaluate CS-TBMEs using the standard matrix elements of the potential in Eq.(\ref{potmat}).
Additionally, it requires the calculation of the radial overlaps of real and CS HO wave functions of the same partial wave:
\begin{equation}\label{overlap}
\!\! \braket{ R_{nl}(r e^{-i\theta}) | R_{n' l}} = \int_0^{+\infty} \!\!\!\!\!\!\!\! R_{nl}(re^{-i\theta}) R_{n' l}(r)~r^2~dr,
\end{equation}
}
which is straightforward to calculate numerically.
Even though convergence is reached with a fairly large number of states in Eq.(\ref{cstbme3}), 
as about 40 HO relative states per partial wave are needed for that matter, the relative CS-TBME of Eq.(\ref{cstbme3}) is very quick to calculate.
In fact, the most expensive part is the Talmi-Brody-Moshinsky expansion generating TBMEs, as in NCSM.

In  numerical calculations, we will use the Daejeon16 interaction, which is of the same form as $\tilde{V}$. 
We note that the separable form of the potential  Daejeon16 is not an approximation as the matrix elements
$ V^{S,S',T}_{\alpha,\beta}$ were fitted to experimental data. 
The CS-TBME of the Coulomb interaction can be directly calculated by the expression of Eq.(\ref{integral}) due to its analytical character.

Three-body forces are crucial for a realistic and predictive understanding of nuclear structure and dynamics. Their inclusion in shell model calculations has significantly improved the agreement with experimental data and provided deeper insights into the nature of nuclear forces. However, implementing these forces poses significant challenges due to computational complexity, the nature of the most adequate effective interactions to be used, and uncertainty in parameterization.
Additionally, there is a specific issue with CS; the existing mathematical framework has only been developed for two-body interactions. Currently, we are not aware of any CS calculations that incorporate a three-body force.

\section{Numerical results}
The deuteron nucleus is especially well suited to test the validity of our CS-NCSM method.
Indeed, as it possesses only two nucleons, the Schr\"odinger equation can be readily handled using relative coordinates, so that CS-NCSM results can be directly compared to the exact deuteron energy.
Added to that, all eigenvalues of the CS-NCSM Hamiltonian matrix can be determined for the deuteron, so that we can compare the numerically determined full CS-NCSM spectrum with the predictions of CS theory. 

\begin{table}\caption{\label{deuteron_table} The deuteron energy (in MeV) evaluated in several $N \hbar \omega$ spaces with NCSM and CS-NCSM.
The parameter of the Lawson correction is fixed at $\beta=1$ and c.m.~spuriosity is about $10^{-9}$ MeV or smaller in all cases.
The exact energy is $-2.224$ MeV.}
\begin{tabular}{|c|c|c|}
\hline \hline
$N$&\multicolumn{1}{c|}{E(NCSM)}&\multicolumn{1}{c|}{E(CS-NCSM)}\\
\hline
6&--2.112&--2.135+i 1.286$\times 10^{-2}$\\
8&--2.147&--2.157--i 1.409$\times 10^{-2}$\\
10&--2.198&--2.207--i 1.542$\times 10^{-2}$\\
12&--2.202&--2.205--i 8.438$\times 10^{-3}$\\
\hline \hline
\end{tabular}
\end{table} 
The numerical results obtained for deuteron energy in NCSM and CS-NCSM are presented in Tab.~\ref{deuteron_table}.
The HO basis energy is fixed at $\hbar\omega = 12.5$ MeV and the Daejeon16 interaction \cite{shi16} is used as nucleon-nucleon potential.
We calculated the expected value of $H_{\textrm{c.m.}}^{\textrm{HO}} - (3/2) \hbar \omega$ (see Eq.(\ref{nocore})), which characterizes c.m.~spuriosity.
It has been noticed to be negligible, being of the order of $10^{-9}$ MeV or smaller, so that the c.m.~part of eigenstates is almost exactly a $0s$ HO state.
One can see that the deuteron energy quickly converges when $N$ increases, as it is only 20 keV above the exact value of $-2.224$ MeV in a 12 $\hbar \omega$ basis model space.
Therefore, both NCSM and CS-NCSM precisely reproduce the deuteron wave function in all considered model spaces.

Theoretically, all imaginary parts of the energies of the CS-NCSM method should be equal to zero, but we obtain around 10 keV in absolute value instead in our calculations.
This deviation is due to the truncation of the many-body Hilbert space, as $6 \leq N \leq 12$ (see Tab.~\ref{deuteron_table}).
Nevertheless, one can observe a slow decrease of imaginary parts in modulus as the basis size increases, so that it can be expected to vanish in the infinite Hilbert space.

\begin{figure}
\includegraphics[width=8.6cm]{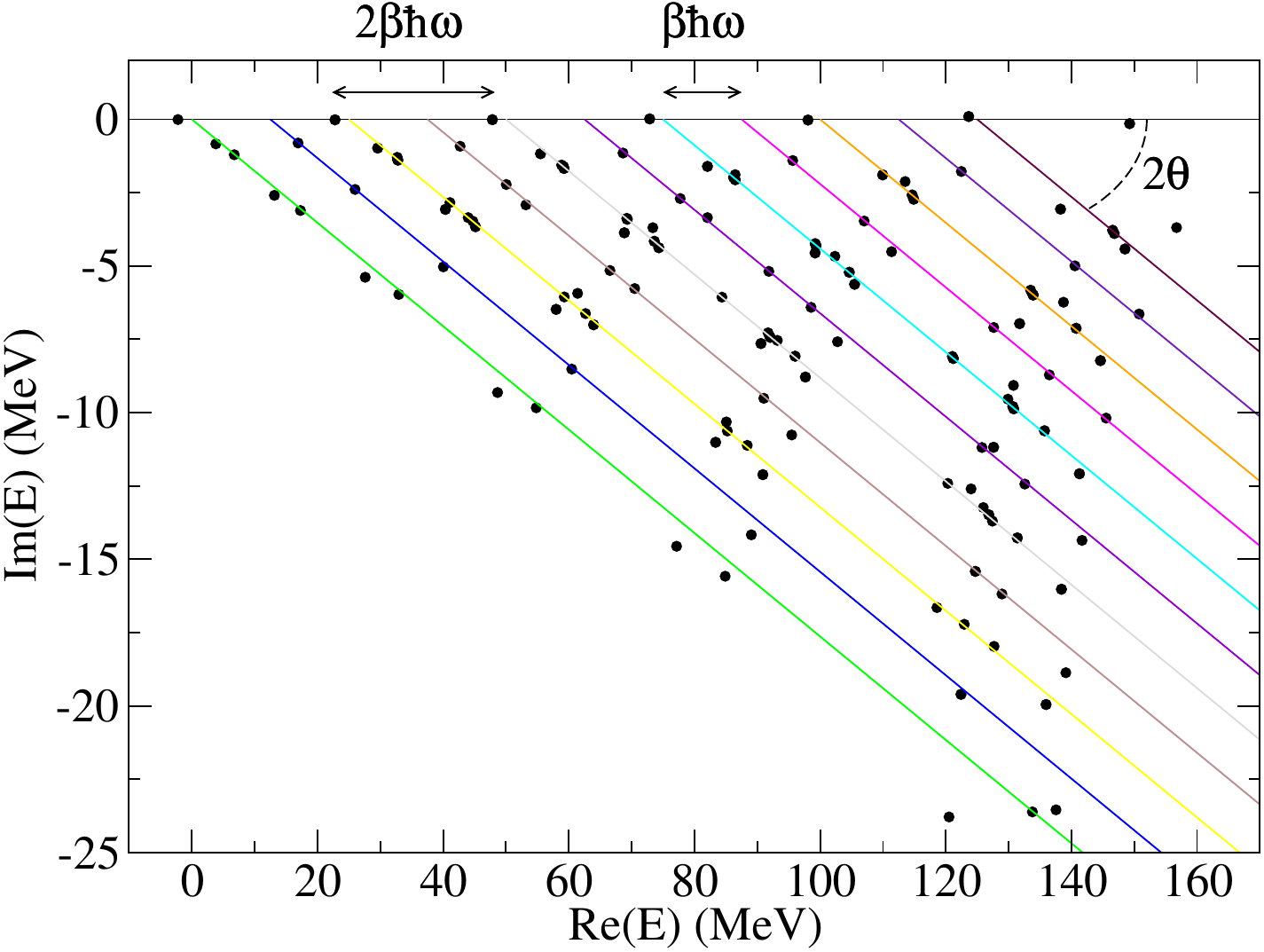}
\caption{\label{energy}
The $1^+$ spectrum of the deuteron nucleus in the CS-NCSM model space of $12 \hbar \omega$ and a CS angle of $\theta=5$ degrees is depicted.
Black dots represent the eigenvalues obtained from the numerical diagonalization of the CS Hamiltonian matrix.
The colored lines correspond to the theoretical continuum spectrum after rotation.
The solid line arc represents the angle of $2\theta$ between the rotated contour and the real axis.
Theoretical shifts of $\hbar\omega$ (contours) and $2\hbar\omega$ (ground states) are depicted (see text for details). }
\centering
\end{figure}
The full deuteron $1^+$ spectrum obtained in the CS-NCSM model is illustrated in Fig.~\ref{energy}, using the same Hamiltonian as above and a $12 \hbar \omega$ model space.
It is apparent that the bound state at $-2.2$ MeV regularly appears every $2 \hbar \omega$ value.
One can also notice several thresholds and rotated-down continuum states in Fig.~\ref{energy}, shifted from each other by $\hbar\omega$. 
The energy shift values of the bound states and rotated continuums are different since the continuum states can have a positive or negative intrinsic parity, whereas the repeated intrinsic ground state always bears $J^\pi = 1^+$.
For scattering states, the c.m.~energy is equal to $(2N + L) \beta \hbar \omega$, $L$ being even or odd, so that it augments by $\beta \hbar \omega$ every time $L$ increases by one unit.
Conversely, as the ground state has its intrinsic and c.m.~parities both positive, the c.m.~orbital angular momentum $L$ must be even, so that the c.m.~energy $(2N + L) \beta \hbar \omega$ increases by $2\beta\hbar\omega$ instead.
The numerical results then clearly align with theoretical predictions (see Fig.~\ref{energy}).
We mention that CS using realistic interactions had already been applied to the two-nucleon system in Refs.\cite{pap15,pap15b}, where relative coordinates were used to represent the proton-neutron system,

Bound state calculations have then been carried out for the bound $^3{\rm H}$,  $^3{\rm He}$ and $^4{\rm He}$ nuclei,
along with the resonance ground states of $^5$He and $^5$Li.
The parameters of the HO basis are the same as in Refs.\cite{shi16,shi18}, whereby $\hbar\omega$ is 12.5 MeV for $^3$H, $^3$He and 17.5 MeV for $^4{\rm He}$, $^5$He, $^5$Li.
The results, calculated in $6 \hbar \omega$, $8 \hbar \omega$, $10 \hbar \omega$ and $12 \hbar \omega$ model spaces, are displayed in Tab.~\ref{E_Gamma_bound_table}.
c.m.~spuriosity has been checked to be negligible in all these cases, as it is of the order of that found in deuteron.
For comparison, we also mention the results of Ref.\cite{shi16} in Tab.~\ref{E_Gamma_bound_table} where a much larger NCSM model space, of $16 \hbar \omega$, is used.

Let firstly consider binding energies in Tab.~\ref{E_Gamma_bound_table}.
One can notice that one has a fairly large dependence on the $\theta$ rotation angle in a $6 \hbar \omega$ model space, as variations of 200-400 keV occur when going from 0 to 10 degrees.
Conversely, the $\theta$ dependence of binding energies is very mild in a $12 \hbar \omega$ model space, as they are of the order of tens of keV, 100 keV being a maximum.
The binding energies of bound states are also very close to the $16 \hbar \omega$ results of Ref.\cite{shi16}.
This shows that the CS-NCSM cannot be utilized in small model spaces for bound states, and that convergence should be basically attained with the considered $N \hbar \omega$ truncation.

The imaginary part of bound state energies is also a good test of the accuracy of the CS-NCSM method, as, being theoretically vanishing, they measure the error due to model space truncation.
On the one hand, in Tab.~\ref{E_Gamma_bound_table} (top), the imaginary part of bound state energies in a $6 \hbar \omega$ model space ranges from 50 keV to 150 keV in absolute value when $\theta$ increases,
which reflects the variation of 200-400 keV of their associated real parts.
On the other hand, in Tab.~\ref{E_Gamma_bound_table} (bottom), the imaginary part of bound state energies in a $12 \hbar \omega$ model space is of the order of 0.5 to 5 keV, which can be deemed as zero numerically. 

\begin{widetext}

\begin{table}
\caption{\label{E_Gamma_bound_table}
The ground state energies in MeV of bound light nuclei in NCSM and CS-NCSM models in $6 \hbar \omega,8 \hbar \omega,10 \hbar \omega,12 \hbar \omega$ model spaces (fourth to last column).
For comparison, we also provide the results of Ref.\cite{shi16} (in brackets), using NCSM with a $16 \hbar \omega$ model space (third column).}
\begin{tabular}{|c|c|c|c|c|c|c|c|}
\hline \hline
& $N (\hbar \omega$) & Ref.\cite{shi16} & NCSM & CS-NCSM (2.5$^\circ$) & CS-NCSM (5$^\circ$) & CS-NCSM (7.5$^\circ$) & CS-NCSM (10$^\circ$) \\
\hline
$^3$H  & 6 & --- &  --8.264 &  --8.280 -- i 0.059 &  --8.326 + i 0.109 &  --8.397 + i 0.139 &  --8.487 + i 0.134 \\
$^3$He & 6 & --- &  --7.565 &  --7.580 +  i 0.054 &  --7.624 + i 0.101 &  --7.691 + i 0.129 &  --7.776 + i 0.124 \\
$^4$He & 6 & --- & --28.311 & --28.319 +  i 0.028 & --28.338 + i 0.050 & --28.347 + i 0.068 & --28.306 + i 0.108 \\
$^3$H  & 8 & --- &  --8.394 &  --8.400 + i 0.013               &  --8.419 + i 0.020               &  --8.446 + i 0.015 &  --8.472 -- i 6.450$\times 10^{-3}$ \\
$^3$He & 8 & --- &  --7.693 &  --7.700 + i 0.011               &  --7.718 + i 0.016               &  --7.744 + i 0.011 &  --7.768 -- i 0.010 \\
$^4$He & 8 & --- & --28.358 & --28.361 + i 5.595$\times 10^{-3}$ & --28.369 + i 9.313$\times 10^{-3}$ & --28.384 + i 0.011 & --28.417 +  i 0.015 \\
$^3$H  & 10 & --- &  --8.428 &  --8.431 + i 5.171$\times 10^{-3}$ &  --8.439 +  i 8.544$\times 10^{-3}$ &  --8.451 +  i 9.354$\times 10^{-3}$ &  --8.463 +  i 9.491$\times 10^{-3}$ \\
$^3$He & 10 & --- &  --7.729 &  --7.732 + i 4.497$\times 10^{-3}$ &  --7.741 +  i 7.467$\times 10^{-3}$ &  --7.753 +  i 8.378$\times 10^{-3}$ &  --7.766 +  i 9.231$\times 10^{-3}$ \\
$^4$He & 10 & --- & --28.370 & --28.371 + i 2.695$\times 10^{-4}$ & --28.374 -- i 9.849$\times 10^{-4}$ & --28.377 -- i 6.277$\times 10^{-3}$ & --28.377 -- i 0.023 \\
$^3$H  & (16)12 &  (--8.442) & --8.436  &  --8.438 + i 5.079$\times 10^{-4}$ &  --8.440 +  i 2.922$\times 10^{-5}$ &  --8.444 -- i 2.235$\times 10^{-3}$ &  --8.447 -- i 7.049$\times 10^{-3}$\\
$^3$He & (16)12 &  (--7.744) & --7.738  &  --7.739 + i 1.650$\times 10^{-4}$ &  --7.742 -- i 5.343$\times 10^{-4}$ &  --7.746 -- i 2.811$\times 10^{-3}$ &  --7.749 -- i 7.422$\times 10^{-3}$\\
$^4$He & (16)12 & (--28.372) & --28.371 & --28.371 + i 6.658$\times 10^{-5}$ & --28.372 +  i 9.526$\times 10^{-5}$ & --28.371 +  i 7.200$\times 10^{-4}$ & --28.366 -- i 5.426$\times 10^{-3}$\\
\hline \hline
\end{tabular}
\end{table}
\end{widetext}

\begin{figure}
\begin{tabular}{c}
\includegraphics[width=0.9\columnwidth]{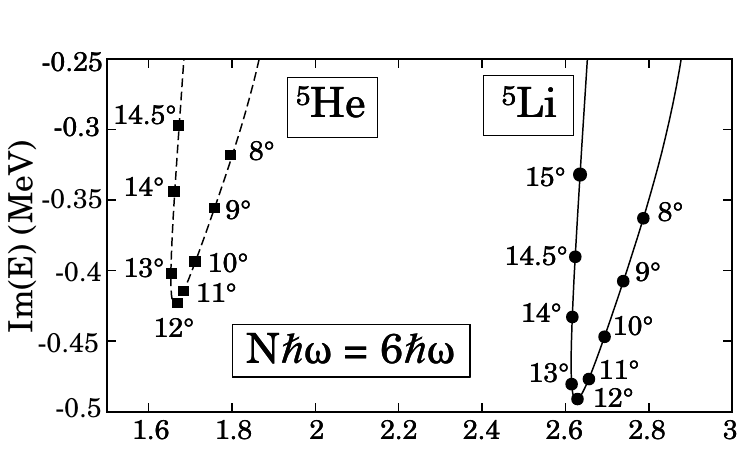} \\
\includegraphics[width=0.9\columnwidth]{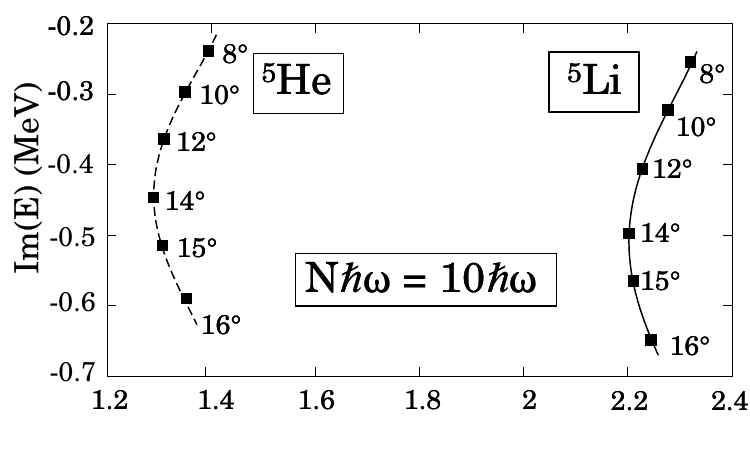} \\
\includegraphics[width=0.9\columnwidth]{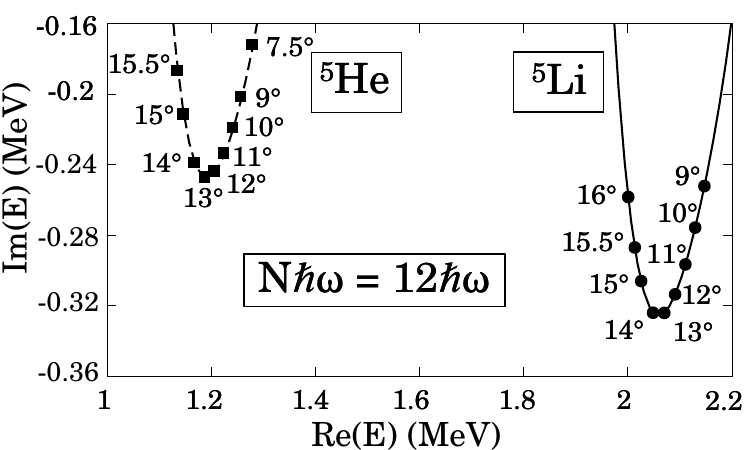}
\centering
\end{tabular}
\caption{\label{E_Gamma_fits}
  Complex eigenenergies of $^5$He (full squares) and $^5$Li (full circles) in the CS-NCSM model using $6 \hbar \omega, 10 \hbar \omega, 12 \hbar \omega$ model spaces (top, middle, bottom, repsectively) as a function of CS angles of $\theta=7.5-16$ degrees.
Energies are provided with respect to $^4$He.
A polynomial fit linking the different complex eigenenergies is illustrated by dashed and solid lines for $^5$He and $^5$Li, respectively.
Results are not provided for the $8 \hbar \omega$ model space because complex energies do not present optima in $\theta$ contours (see text for details).}
\end{figure}

The ground states of $^5$He and $^5$Li are unbound, so that their eigenenergies are complex, with their imaginary part directly providing their particle-emission widths (see Fig.~\ref{E_Gamma_fits}).

In both CS and GSM methods, the non-Hermitian Hamiltonian yields complex eigenvalues.
The $\theta$ trajectory method is a reliable tool for identifying resonant states and extracting their parameters \cite{moi98,myo14b,zhu22}.
This approach leverages CS theory, particularly the rotation of the continuum and the $\theta$ independence of bound and resonance poles.
The method tracks the evolution of the discrete eigenvalues of the CS Hamiltonian with varying scaling angle $\theta$.
In the complex energy plane, bound-state eigenvalues remain nearly invariant with $\theta$, while resonance eigenvalues show $\theta$-dependent trajectories that stabilize at specific complex energy values \cite{mas14,mic21}.
This stabilization identifies the resonance.
By analyzing these trajectories, one can determine the optimal $\theta$ where resonance eigenvalues are most stable, ensuring an accurate extraction of resonance parameters.
We also investigated c.m.~spuriosity in $^5$He and $^5$Li and noticed that it amounts to about 10 keV in their ground states.
While it is larger than those of previously studied bound states, it is sufficiently small to ensure that c.m.~excitations are physically removed.

\begin{table}
\caption{\label{opt_E_Gamma_5He_5Li}
The optimal values of the separation energy (in MeV) and width (in keV) of the ground states of $^5$He and $^5$Li provided by the fit of $\theta$-dependent complex energies (see Fig.~\ref{E_Gamma_fits})
in the CS-NCSM $6 \hbar \omega, 10 \hbar \omega, 12 \hbar \omega$ model spaces (denoted as 6,10,12 as subscripts below).
The values purported in Ref.\cite{shi18} and obtained with the Daejeon16 interaction by employing the SS-HORSE method (hereby abbreviated by the subscript SH) are also provided for comparison.
Results are not provided for the $8 \hbar \omega$ model space because complex energies do not present optima in $\theta$ contours (see text for details).}
\begin{tabular}{|c|c|c|c|c|c|c|c|c|}
\hline \hline
& $E_{6}$ & $\Gamma_{6}$ & $E_{10}$ & $\Gamma_{10}$ & $E_{12}$ & $\Gamma_{12}$ & $E_{\rm SH}$ & $\Gamma_{\rm SH}$ \\
\hline
$^5$He & 1.663 & 847 & 1.290 & 866  & 1.192 & 495 & 0.68  & 520   \\
$^5$Li & 2.627 & 983 & 2.190 & 1050 & 2.059 & 651 & 1.052 & 1050 \\
\hline \hline
\end{tabular}
\end{table}
The $\theta$ trajectories are displayed on Fig.~\ref{E_Gamma_fits} via the use of a polynomial fit applied to the $^5$He and $^5$Li ground states complex energies.
The complex-energy extrema clearly appear, which provide the optimal values of energy and width of the ground states of $^5$He and $^5$Li (see Tab.~\ref{opt_E_Gamma_5He_5Li}).
As in Refs.\cite{mas14,kru90,mic21}, the extrema appear at $\theta \simeq 13^\circ$.

Conversely, our separation energy values differ from those obtained in the SS-HORSE model of Ref.\cite{shi18} (see Tab.~\ref{opt_E_Gamma_5He_5Li}), as the CS-NCSM energies are 500 keV to 1 MeV too bound, comparatively.
While our neutron-emission width for $^5$He is very close to that of Ref.\cite{shi18}, as both lie around 500 keV, our proton-emission width for $^5$Li is 350 keV smaller than that provided by the SS-HORSE method.
Nevertheless, we do not deem this difference as problematic because the method utilized in Ref.\cite{shi18} to identify resonances makes use of scattering phase shifts, 
whereas that of GSM and CS-NCSM relies on the generalized variational principle.
Unless one considers narrow widths, typically less than 100 keV, these two methods provide different values, whose discrepancy increases along with resonance widths \cite{muk10}.
This is, in fact, the case in our examples, as widths are around 500 keV for both $^5$He and $^5$Li in a $12 \hbar \omega$ model space.

The $3/2^-$ ground states of $^5$He and $^5$Li have been calculated with NCGSM with the Daejeon16 interaction,
whereby the $p_{3/2}$ partial wave (neutron for $^5$He, proton for $^5$Li), is represented by the Berggren basis and is discretized with 15 states.
All other one-body states are of HO character, so that the many-body basis is akin to a $6 \hbar \omega$ model space.
A maximum of four nucleons in the continuum is also imposed.
It is not possible in practice to extend model spaces beyond that limit, as calculations quickly become unstable otherwise.
We also verified beforehand that the energy of $^4$He is -28.315 MeV, which is virtually the same value as that provided by NCSM.
The obtained excitation energies of the ground states of $^5$He and $^5$Li are about 1.4 MeV and 2.2 MeV, which are only about 200-400 keV smaller than those of CS-NCSM, respectively.
Nevertheless, the associated widths are about 60 keV and 370 keV for $^5$He and $^5$Li, instead of about 500 keV and 650 keV in CS-NCSM using a $12 \hbar \omega$ model space, respectively.
That discrepancy is surely due to the strong dependence of widths on excitation energies (see Tab.~\ref{opt_E_Gamma_5He_5Li}).
Indeed, the model space of the NCGSM calculation, possessing a complete $p_{3/2}$ partial wave, is effectively larger than that of the CS-NCSM in a $12 \hbar \omega$ model space.
Comparatively, the obtained energy and width for $^5$He in a $12 \hbar \omega$ model space are closer to those previously calculated with NCGSM in Ref.\cite{pap13},
where they are respectively equal to 1.17 MeV and 400 keV.
This better agreement was obtained probably because the model space of Ref.\cite{pap13} is smaller than that of the present NCGSM calculations,
as only a few HO $dfg$ shells were added to the Berggren basis of $sp$ contours therein.

Interestingly, the energies and widths of $^5$He and $^5$Li computed in a $6 \hbar \omega$ model space are qualitatively correct (see Fig.~\ref{E_Gamma_fits}),
as energies are about 500 keV too unbound and widths only differ by about 300-400 keV, that compared to the more precise calculation performed with a $12 \hbar \omega$ model space.
This suggests that the CS-NCSM can be relatively accurate for unbound states even in rather small model spaces.

Conversely, we could not reach similar results in the $8 \hbar \omega$ model space.
No optimum for $A=5$ complex energies could be found in the $8 \hbar \omega$ space, as energy and width continuously increases with the $\theta$ angle.
Correspondingly, the energy of $^4$He augments along with $\theta$, even though it remains close to its NCSM value when $\theta \sim 0$ (see Tab.~\ref{E_Gamma_bound_table}).
As a consequence, the $8 \hbar \omega$ model space is not sufficient to provide accurate complex energies in $A=5$ nuclei, whereas it is the case in the $6 \hbar \omega$ model space.
It is thus very likely that accidental cancellations occur in the $6 \hbar \omega$ model space and not in the $8 \hbar \omega$ model space,
as the complex energies implemented with the $6 \hbar \omega$ and $12 \hbar \omega$ model spaces behave the same qualitatively.

While the separation energies and widths of $^5$He and $^5$Li are well defined and physical in the $10 \hbar \omega$ model space,
the $\theta$ trajectory bears a different shape compared to that encountered in the $6 \hbar \omega$ and $12 \hbar \omega$ model spaces.
Indeed, energy increases along $\theta$, while width exhibits a minimum in the $6 \hbar \omega$ and $12 \hbar \omega$ model spaces for both $^5$He and $^5$Li ground states,
while it is the opposite in the $10 \hbar \omega$ model space.
This does not contradict the ABC theorem, as the $\theta$ trajectory is not constrained by the considered Hamiltonian \cite{moi98,myo14b,zhu22}.
Indeed, the generalized variational principle pertaining to the $\theta$ trajectory can provide a saddle point in the complex plane, which is not necessarily a minimum for energy.
Added to that, the energies and widths derived from the $\theta$ stabilization method diminish while model space dimension augments, which is expected from the generalized variational principle.
Note that the different shapes encountered in $\theta$ trajectories had already been discussed in Ref.\cite{moi98}.

The main issue to solve for that matter is the value of the maximal width that can be handled in complex-scaled no-core shell model.
Indeed, $\Gamma \sim 1$ MeV is about the current limit that could be managed at present.
While most resonances in light nuclei bear a width smaller or comparable, it is not the case in the spectrum of $^4$He, for example, where widths typically reach 5 MeV or more.
This would demand to be able to tackle Hamiltonians generated by angles between 15 and 20 degrees, which is not possible in practice for the moment due to numerical instability.
As model space dimensions of nuclei in the $A = 5-10$ region are tractable in the context of CS-NCSM, this issue must be solved in priority in complex-scaled no-core shell model for the study of unbound nuclear states.

\section{Conclusion}
The complex scaling method is an efficient way to determine resonance states as it transforms the Hamiltonian operator so that resonance eigenstates become square-integrable.
This allows for the calculation of resonance states using bound state type methods.
However, complex scaling can be used only if the Hamiltonian operator is defined with relative coordinates, which has probably prevented its extension to nuclear shell model, as the latter uses single-particle laboratory coordinates.
It has been shown that by applying complex scaling only to the intrinsic part of the Hamiltonian operator of the no-core shell model, a model suitable for simultaneously determining bound and resonance states can be obtained.

Numerical calculations have been performed for systems consisting of a few nucleons, and our ground state and resonance energies align well with previously published results.
We have shown that the CS-NCSM possesses a low numerical cost compared to the other methods dedicated to the study of weakly bound and unbound nuclei.
Added to that, we could evaluate the particle-emission widths of unbound nuclei within our complex-scaled no-core shell model using the generalized variational principle applied to complex-scaling angle $\theta$.
Indeed, the dependence of eigenergies on the complex-scaling angle $\theta$ is rather large, so that a stabilization method, based on finding the complex energy stationary with respect of small changes of $\theta$, 
must be performed to identify unbound states precisely.
The overall study is thus promisive, so that it is believed that the suggested complex-scaled no-core shell model method will become valuable for studying low-energy nuclear resonances in a near future.

\textit{Acknowledgments} ---
This work has been supported by the National Key R\&D Program of China under Grant No.~2023YFA1606403 and 2024YFE0109802;
the CAS President’s International Fellowship Initiative (PIFI) No.~2024PVA0109;
the Youth Innovation Promotion Association of Chinese Academy of Sciences No.~Y2021414;
the National Natural Science Foundation of China Nos.~12205340, 12347106, and 12121005;
the Gansu Natural Science Foundation under Grant No.~22JR5RA123 and 23JRRA614;
the Key Research Program of the Chinese Academy of Sciences under Grant No.~XDPB15;
the National Natural Science Foundation of China under Grant No.~12375117;
the State Key Laboratory of Nuclear Physics and Technology, Peking University under Grant No.~NPT2020KFY13.
A.~T.~Kruppa thanks the CAS PIFI scholarship for support.

\section*{References}

\bibliography{references}

\end{document}